\documentclass[lettersize,journal]{IEEEtran}
\usepackage{amsmath,amsfonts}
\usepackage{algorithmic}
\usepackage{algorithm}
\usepackage{array}
\usepackage[caption=false,font=normalsize,labelfont=normalsize,textfont=sf]{subfig}
\usepackage{textcomp}
\usepackage{stfloats}
\usepackage{url}
\usepackage{verbatim}
\usepackage{graphicx}
\usepackage{cite}
\begin{document}

\title{Integrating Satellites and Mobile Edge Computing for 6G Wide-Area Edge Intelligence: Minimal Structures and Systematic Thinking}

\author{Yueshan Lin, Wei Feng, \IEEEmembership{Senior Member, IEEE}, Ting Zhou, \IEEEmembership{Member, IEEE}, Yanmin Wang, Yunfei Chen, \IEEEmembership{Senior Member, IEEE}, Ning Ge, \IEEEmembership{Member, IEEE}, and Cheng-Xiang~Wang, \IEEEmembership{Fellow, IEEE}
\thanks{Y. Lin, W. Feng, and N. Ge are with the Department of Electronic Engineering, Tsinghua University, Beijing 100084, China (email: lin-ys17@tsinghua.org.cn, fengwei@tsinghua.edu.cn, gening@tsinghua.edu.cn).}
\thanks{T. Zhou is with the Shanghai Advanced Research Institute, Chinese Academy of Sciences, Shanghai 201210, China (email: zhouting@sari.ac.cn).}
\thanks{Y.~Wang is with the School of Information Engineering, Minzu University of China, Beijing 100041, China (email: yanmin-226@163.com).}
\thanks{Y. Chen is with the School of Engineering, University of Warwick, Coventry CV4 7AL, U.K. (email: Yunfei.Chen@warwick.ac.uk).}
\thanks{C.-X. Wang is with the National Mobile Communications Research Laboratory, School of Information Science and Engineering, Southeast University, Nanjing 210096, China, and also with Purple Mountain Laboratories, Nanjing 211111, China (email: chxwang@seu.edu.cn).}

}



\maketitle

\begin{abstract}
The sixth-generation (6G) network will shift its focus to supporting everything including various machine-type devices (MTDs) in an everyone-centric manner. To ubiquitously cover the MTDs working in rural and disastrous areas, satellite communications become indispensable, while mobile edge computing (MEC) also plays an increasingly crucial role. Their sophisticated integration enables wide-area edge intelligence which promises to facilitate globally-distributed customized services. In this article, we present typical use cases of integrated satellite-MEC networks and discuss the main challenges therein. Inspired by the protein structure and the systematic engineering methodology, we propose three minimal integrating structures, based on which a complex integrated satellite-MEC network can be treated as their extension and combination. We discuss the unique characteristics and key problems of each minimal structure. Accordingly, we establish an on-demand network orchestration framework to enrich the hierarchy of network management, which further leads to a process-oriented network optimization method. On that basis, a case study is utilized to showcase the benefits of on-demand network orchestration and process-oriented network optimization. Finally, we outline potential research issues to envision a more intelligent, more secure, and greener integrated network.
\end{abstract}


\section{Introduction}
In the future, an increasing number of wireless sensors, industrial robots, and intelligent machines will be deployed to free human beings. This stimulates a focus shift of the future sixth-generation (6G) network from serving human beings to supporting various intelligent machine-type devices (MTDs). These MTDs may be employed in disastrous areas for emergency rescue ({\it e.g.}, vital sign detection). They can also be deployed in remote areas, such as oceans, deserts and forests, for environmental monitoring and resource exploitation. In these harsh areas, constructing terrestrial cellular networks, as in the fifth-generation (5G) system, will be difficult or extremely expensive, due to damaged infrastructures and tough geographical conditions. Instead, a non-terrestrial network via satellites and unmanned aerial vehicles (UAVs), may be used to fill the coverage gap and offer everyone-centric customized services \cite{intro 03}. 

Despite the advantage of global coverage \cite{intro 01}, satellite communications also face their inherent challenges in terms of limited data rate and large latency, which pose difficulty in satisfying the MTDs' service requirements. For example, some MTDs may have to use cloud computing for data processing, due to their limited computing and energy resources on board. However, the bandwidth resources of satellites are limited, and offloading massive data via satellites leads to an unacceptable delay. Furthermore, some MTDs may participate in a time-critical activity, which requires delay-sensitive services. All these issues pose challenges to the use of satellite communications for MTDs, and will inevitably lead to poor user satisfaction ratios \cite{intro 03}.

To meet the service demands of these MTDs in an everyone-centric manner \cite{intro 03}, edge intelligence is required to replace traditional central cloud computing. Specifically, by enabling artificial intelligence (AI) applications at the network edge, MTDs are endowed with low-latency data processing and decision making capabilities. Mobile edge computing (MEC) is crucial for empowering this edge intelligence paradigm, which may further evolve into a more promising network AI architecture \cite{intro 03}. Current 5G networks have partially used MEC, which shows the benefit of integrating cellular communications and MEC. Thus, it is envisioned that integrating satellites and MEC can support better MTDs in the aforementioned disastrous and remote areas \cite{intro 04}\cite{intro 05}. However, the integration of MEC and satellite networks remains an open issue.

In this article, we will provide typical use cases of the integrated satellite-MEC network. The main challenges of satellite-MEC integration will then be discussed. We will use systematic thinking to address these challenges. Particularly, we will regard the network as the extension and combination of several minimal integrating structures, whose characteristics and key problems will be discussed. Moreover, an on-demand network orchestration framework will be designed to introduce a medium timescale for network adjustment. A process-oriented optimization method will be accordingly proposed under this framework to improve the network resource efficiency and offer wide-area customized services. Then, a case study on process-oriented optimization in an extended minimal structure will be presented. Several future research directions of the integrated satellite-MEC network will finally be outlined. 

\begin{figure*}[tb]
	\centering
	{\includegraphics[width=6.5in]{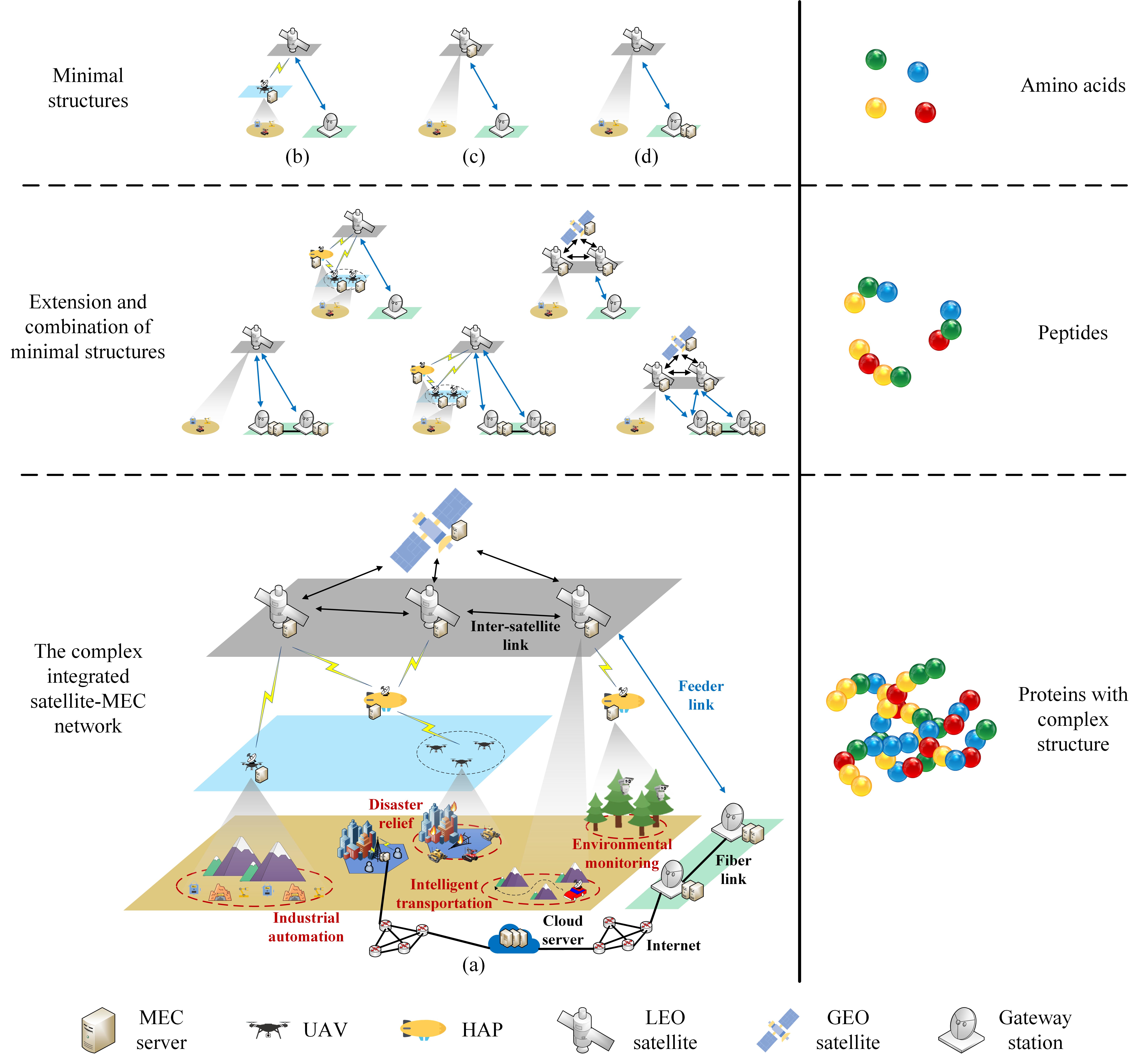}%
		\label{system model}}
	\caption{A systematic view of the integrated satellite-MEC network (left) inspired by the protein structure (right).}
	\label{fig_sim}
\end{figure*}

\section{Use Cases of Satellite-MEC Integration}
The integrated satellite-MEC network is envisioned to support various MTDs in an everyone-centric manner \cite{intro 03}. As shown in Fig. 1(a), we summarize typical use cases in the following.
\IEEEpubidadjcol
\begin{itemize}
	\item{{\bf Intelligent environmental monitoring}. In this use case, a large number of fine-grained sensors are widely distributed in remote areas to acquire environmental data. A massive amount of data need to be transmitted and analyzed to detect environmental anomalies (floods, water pollution, forest fire, {\it etc.}), which can consume excessive communication resources. As a solution, the integrated satellite-MEC network is enabled to provide wide-area coverage for the sensors to allow the massive data to be processed with close proximity to users \cite{use case 01}}. 
	\item{{\bf Efficient disaster relief}. Natural or man-made disasters (earthquakes, floods, fires, explosions, {\it etc.}) require that both rescue groups from outside and first responders ({\it e.g.}, vital sign detectors) from inside take immediate action. However, terrestrial communication infrastructures in disastrous areas are often damaged. When disasters occur, the integrated satellite-MEC network can be rapidly deployed and provide situation information for rescue groups. Meanwhile, they can also assist first responders with real-time decision-making through local computation.}
	\item{{\bf Intelligent transportation}. This use case provides vehicles with diverse services, such as automatic control assistance and dynamic path planning. For remote vehicles ({\it e.g.}, ocean vehicles) and highly mobile vehicles, terrestrial networks are not applicable. The integrated satellite-MEC network can provide ubiquitous coverage for these vehicles. Moreover, they can process the massive data collected by multiple vehicles in real time, which further enables automatic control assistance and dynamic path planning services \cite{use case 02}}.
	\item{{\bf Industrial automation}. The exploration and acquisition of industrial materials often take place in rural or remote areas, where multiple sensors and actuators execute their tasks cooperatively. Industrial automation requires that the collected sensor data should be processed in real time to provide instructions for actuators, forming a closed-loop control system. To meet such requirements, the integrated satellite-MEC network can provide communication coverage and real-time computing services for these sensors and actuators.}
\end{itemize}

\section{Challenges of the Integrated Satellite-MEC Network}
\subsection{Unique Characteristics of Service Requirements}
In terms of achieving edge intelligence, it is widely acknowledged that a complementary relationship between terrestrial-MEC and satellite-MEC networks is necessary. In contrast to the terrestrial network, the integrated satellite-MEC network mainly aims at application scenarios in remote or disastrous areas, as discussed before. The service requirements in these application scenarios have these unique characteristics. First, MTDs are expected to be dispersed over extensively wide geographical areas. The spatial distribution of their service requirements is remarkably sparser than that in the terrestrial network. A typical example is the buoys that collect hydrological information for ocean monitoring \cite{use case 01}. Moreover, the temporal and spatial distributions of the service requirements can be heterogeneous and highly dynamic. For instance, industrial automation and efficient disaster relief applications require aggregation of MTDs, and the positions of these aggregates vary over time. These unique characteristics render the system design of terrestrial networks inappropriate due to the low resource efficiency. Therefore, new challenges concerning the integrated satellite-MEC networks' system designs require careful consideration.

\subsection{Limited Communication and Computing Capability}
In addition to the service requirement characteristics, the inherent limitation in communication and computing capability also poses challenges to the integrated satellite-MEC network. On the one hand, the communication capability of the satellite network is inherently limited. The transmission rate of satellite-ground links is relatively low due to the long propagation distance and limited bandwidth resources. In addition, the number of satellites in the network is restricted. This is because of the high costs of space launch and the limited orbit resources. Specifically, although a single low earth orbit (LEO) satellite is expected to have a similar level of throughput as a 5G base station, the number of satellites in a constellation is remarkably smaller \cite{challenge 02}. This makes it difficult for the satellites to satisfy the global service demands. On the other hand, due to the limited capability of satellite communication, MEC servers on satellites and aerial platforms (APs) are necessary in the integrated network for delay-sensitive services. These MEC servers, especially the satellite servers, are under stringent restrictions in terms of size, weight, and power. Moreover, because of the severe electromagnetic radiation in space, the satellite servers need to be radiation-hardened \cite{challenge 03}. This requires special design and incurs extra costs. These factors account for the limited computing capability in integrated satellite-MEC networks.

\begin{figure}[tb]
	\centering
	\includegraphics[width=3.5in]{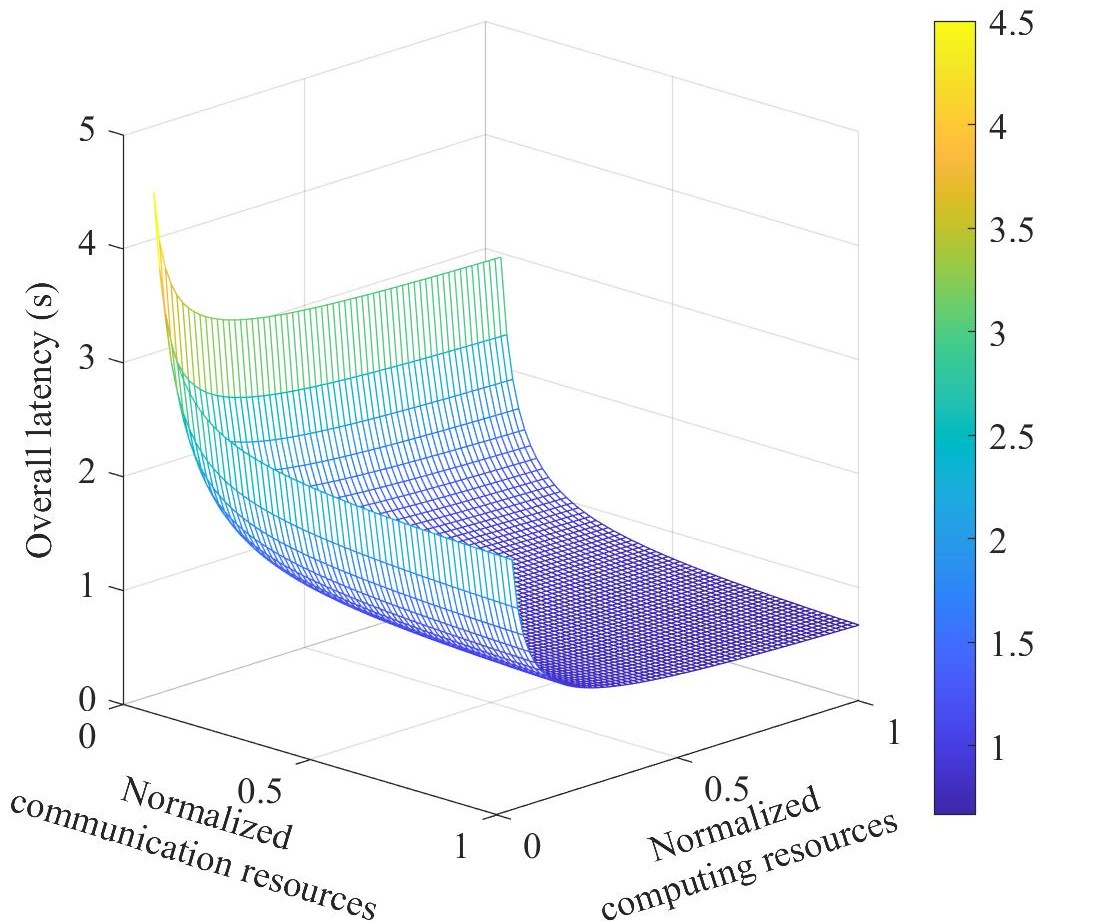}
	\caption{A schematic diagram of the overall latency of a single offloading task with different communication and computing resources provided.}
	\label{resource configuration}
\end{figure}

\subsection{Complexity of Matching Network Resources with Service Requirements}
Fig. 2 shows a schematic diagram of the overall latency of a single offloading task, varying with the communication and computing resources provided. The resources are normalized, and the latency is obtained by linearly combining the reverse of both resources\footnote{In different use cases, the specific relationship between the overall latency and the network resources can be different.}. We observe that the latency decreases quite slowly with both resources in the resource-adequate scenario. Therefore, configuring proper resources to match the service requirements of an offloading task can considerably improve resource efficiency. For a practical system of multiple offloading tasks, such a resource configuration evolves to a complex high-dimensional backpack problem. For the integrated satellite-MEC network, the hierarchical communication and computing resources, such as the diverse transmission links and the multiple tiers of MEC servers, render it difficult to match the resources with the service demands. Moreover, LEO satellites and aerial platforms are highly mobile in the network, which makes the network resources vary over time. The resource configuration problem becomes even more complex when the time dimension is accounted for.

\section{Minimal Integrating Structures}
It is an important but challenging problem to efficiently utilize the limited network resources to meet the wide-area and sparse service demands. To handle this problem, we provide a novel systematic perspective inspired by the protein structure, as shown in Fig. 1. Specifically, we propose three minimal integrating structures and discuss their corresponding characteristics and problems based on the location of MEC server. A more complex integrated satellite-MEC network can be considered as an orchestration of these structures. By figuring out the minimal integrating structures, as well as their extensions and combinations, we can obtain better insight into the complex network.

\subsection{Computing-in-Forward-Link Structure}
As shown in Fig. 1(b), this minimal integrating structure consists of an AP equipped with an MEC server, a satellite, a gateway and multiple MTDs. The AP can be a UAV or a high altitude platform (HAP). With MEC deployment, the AP provides local computation services for MTDs. Due to the limitations of APs in terms of size and energy, the central processing unit (CPU) capability of an on-board MEC server is restricted. In addition, the AP can also relay the task data to the satellite, and the task data are further transmitted to a cloud server for computation. This minimal structure can be extended by considering multiple APs.

Compared to satellites, APs can provide high-speed communications, which leads to lower transmission delay. However, the coverage area of an AP is smaller than that of a satellite. From these perspectives, it can be deduced that the computing-in-forward-link structure is suitable for local-area computation services with ultra-low latency requirements. Examples of these services include robotic machine controls in industrial automation and real-time decision making in efficient disaster relief.

For such a minimal structure and its extensions, multiple challenges exist in terms of system design. First, it is of great importance to determine the on-board CPU capability, as well as the capacities of the satellite-AP link and the AP-user link. It should be noted that the capacities of the two links need to achieve a proper fit to improve resource efficiency. Moreover, in the multi-AP case, transmission links might exist among the APs, which further adds to the difficulty of configuring communication and computing resources.

Another challenge is that proper user association plays an important role in the system design of the multi-AP structure. One typical user association scheme is that each AP is equipped with an MEC server, and each user is associated with one AP. Ding {\it et al.} \cite{model 01 018} considered such a scheme and investigated the joint optimization problem of user association, resource allocation, offloading assignment, and transmit precoding. In addition, we can also assume that each user has access to multiple APs and their on-board servers, forming a coordinated multi-point network. Moreover, some of the APs may not carry MEC servers and only serve as relay nodes.

Due to the mobility of APs, the trajectory of APs and the maximum coverage time also require careful consideration in system design.

\subsection{Computing-on-Orbit Structure}
This minimal integrating structure is composed of a satellite equipped with an MEC server, a gateway and multiple MTDs, as shown in Fig. 1(c). The MTDs' computation tasks can be directly executed at the satellite or further uploaded for cloud computing. The satellite is highly limited in terms of size, weight and energy supply, which places stringent restrictions on the CPU capability of the on-board MEC server. There are several derivatives from this minimal structure. For instance, the space segment can be a constellation of LEO satellites. In addition, a geostationary earth orbit (GEO) satellite and a LEO constellation can coordinately provide edge computing services.

Compared with APs, satellites can provide a much larger coverage to serve more devices. In addition, satellites are in proximity to ground MTDs topologically compared with gateway stations. Therefore, the computing-on-orbit structure can well serve the requirements of computation tasks that span a wide area and require relatively low latency. For instance, path planning for large-area vehicles can utilize this structure to execute their computation tasks. 

Multiple challenges need to be handled for this minimal structure and its extended structures. First, it is not as easy to place MEC servers on satellites as on UAVs or at gateways. On the one hand, to implement edge computing on orbit, on-board processing (OBP) capabilities ({\it e.g.}, modulation/demodulation, encoding/decoding) are necessary on satellites. On the other hand, the severe electromagnetic radiation in space can induce bit-flips and even cause damage to the satellite payloads. This requires that the servers and the OBP modules should be radiation-hardened \cite{challenge 03}. These factors add to the complexity of system design, especially when the limited size, weight and energy of satellites are considered.

With multiple satellites, the placement method of MEC servers is another important challenge. For instance, we consider the scenario of a LEO constellation providing services. Radiation-hardened MEC servers can be placed on all LEO satellites, which leads to high costs of MEC hardening. Another scheme is that the servers are placed on part of the satellites, while the other satellites merely relay data through inter-satellite links (ISLs). Adopting this scheme can reduce the MEC hardening costs, but the ISL costs will be higher. For the MEC server placement problem, this tradeoff requires careful consideration.

In the extended structure where the space segment is a LEO constellation, there are often multiple satellites in the MTD's line of sight. For more computation-intensive applications, it is an intriguing problem to achieve simultaneous offloading to multiple satellites. Song {\it et al.} \cite{model 02 008} investigated this problem and jointly optimized the task division and the transmission power to each satellite. From another angle, Wang {\it et al.} \cite{model 03 001} proposed a game-theoretic approach to optimize the offloading strategy of multiple users.

Finally, LEO satellites are highly mobile and thus their coverage time can be quite limited. Therefore, handover of MEC servers may be needed, which is also a problem worth investigating.

\subsection{Computing-after-Feeder-Link Structure}
As shown in Fig. 1(d), this minimal integrating structure consists of a satellite, a gateway equipped with an MEC server and multiple MTDs. MTDs offload their computation tasks to the gateway through the satellite relay. The computation tasks are executed in the gateway server or they are transmitted to the remote cloud. It should be noted that the MEC server in this minimal structure possesses higher CPU capability compared with the above two minimal structures, since the gateway station allows a larger server size and higher energy consumption. The minimal structure can be extended by considering multiple gateways.

With satellites as relays, the MEC server deployed at the gateway station can meet the demands of large-area computation services. In addition, the CPU capability of the MEC server is higher. However, the communication latency is relatively high since the data may traverse several ISLs before arriving at the gateway. Therefore, this minimum structure is most suitable for wide-area computation-intensive but delay-tolerant services, such as collecting and processing massive sensor data in environmental monitoring.

The computing-after-feeder-link structure and its extensions also raise problems in terms of system design. For instance, the MEC server placement problem is an important challenge. It is typically assumed that multiple gateways work separately. In this case, the CPU capability of each gateway server needs to be properly configured. In addition, the computing resource allocation of each server requires optimization. For instance, Zhang {\it et al.} \cite{model 05 011} proposed a game-theoretic scheme to solve the joint offloading decision and resource allocation problem. Cui {\it et al.} \cite{model 06 066} jointly optimized user association, task scheduling, and resource allocation. Another assumption is that the gateways are interconnected through high-speed fiber links. In this case, the transmission latency among gateways can be ignored, and the problem of server placement degenerates to deciding the total CPU capability of all servers.

\section{On-demand Network Orchestration}
\subsection{Medium-Timescale Network Orchestration}
On-demand adjustments of the integrated satellite-MEC network are of great importance for providing wide-area customized services \cite{intro 03}. As illustrated before, the temporal and spatial distributions of MTD users are sparse. In addition, the communication and computing capability of the integrated network are rather limited. Through dynamically adjusting the network to match the changing service requirements, the resource efficiency can be substantially improved. Because of this, the network can fully utilize its limited resources to provide ubiquitous services.

The current 5G network is mainly adjusted at two different timescales. First, mobile network operators conduct network planning at the timescale of months or years. Network planning adjusts communication and computing infrastructures, such as the number of satellites and the CPU capability of satellite servers. The second is the timescale of milliseconds or less, which is close to the channel coherence time. At such a timescale, the transmission parameters are adjusted, such as channel estimation and beamforming.

However, the service demands often change at a medium timescale ({\it e.g.}, minutes or hours), in terms of number, spatial distribution and service type. Both network adjustments in the 5G network are incapable of matching these changes. Therefore, novel medium-timescale network adjustments need to be introduced in the integrated satellite-MEC network. Moreover, the network can be viewed as a nonlinear orchestration of the three minimal structures, as illustrated above. Therefore, we will establish an on-demand network orchestration framework, which adjusts the minimal structure orchestration at medium timescales to match the varying service requirements. Specifically, this can be achieved by adjusting the activation ratios of certain transmission links and MEC servers for instance. Table 1 summarizes the network adjustments, as well as their corresponding timescales and adjusted parameters.

\begin{table}[t]
	\caption{Comparison of Network Adjustments at Different Timescales}
	\centering
	\renewcommand{\arraystretch}{2}
	\begin{tabular}{|m{0.7in}<{\raggedright}|m{0.7in}<{\raggedright}|m{1.35in}<{\raggedright}|}
		\hline
		& \textbf{Timescale} & \textbf{Adjusted parameter} \\ \hline
		\textbf{Network planning} & months/years & Number of satellites, number of gateways, satellite-user link capacity, ISL capacity, on-board CPU capability \\ \hline
		\textbf{Minimal structure orchestration} & minutes/hours & Activation ratios of servers, activation ratios of ISLs, spot beam management, deployment of UAVs \\ \hline
		\textbf{Adaptive transmissions} & milliseconds & Encoding/decoding, modulation/demodulation, channel estimation, beamforming \\ \hline
	\end{tabular}
\end{table}

Fig. 3 depicts an instance of medium-timescale minimal structure orchestration for four constant time periods. As shown in Fig. 3(a), multiple environmental sensors require updating their data for analysis in the first period. In this case, the activation ratios of gateway servers are high, while the satellite servers are turned off to save energy. Such a network adopts the computing-after-feeder-link structure. The second period sees a decrease in the number of transmitting sensors, as shown in Fig. 3(b). Therefore, the activation ratios of gateway servers are lowered. Fig. 3(c) shows that some mineral occurrences start exploiting resources in the third period. To provide industrial automation services, some satellite servers are turned on, and a UAV is dispatched for delay-sensitive computation offloading. The network is then adjusted to a combination of the computing-in-forward-link structure and the computing-on-orbit structure. In the fourth period, the activation ratios of UAV servers and satellite servers are further raised to adapt to the increase in service demands, as depicted in Fig. 3(d).

\subsection{Process-Oriented Network Optimization}
For medium-timescale network orchestration, some network parameters are adjusted over different time periods, which matches the change of service requirements. However, these parameters, such as the MEC activation ratio in Fig. 3, remain fixed during a time period. This requires that the parameters should be optimally designed to maximize the system performance of the entire period. To address this problem, we propose a process-oriented optimization method, where the process refers to a medium-timescale period. The design of this type of method is challenging, since the accurate process information can be difficult, if not impossible, to acquire. For instance, the process duration is much larger than the channel coherence time. Therefore, the full channel state information (CSI) during the process is considered random, and thus cannot be obtained accurately. To address this challenge, one promising direction is to introduce external information to assist the optimization. For instance, the large-scale CSI of the whole process can be conveniently acquired from a pre-established database, referred to as a {\it radio map} \cite{adjust 01}. Process-oriented optimization can be conducted based on the large-scale CSI. Likewise, we also need a large-scale model that characterizes the service requirements, which is worth further investigation.

\begin{figure*}[t]
	\centering
	\subfloat[]{\includegraphics[width=2.55in]{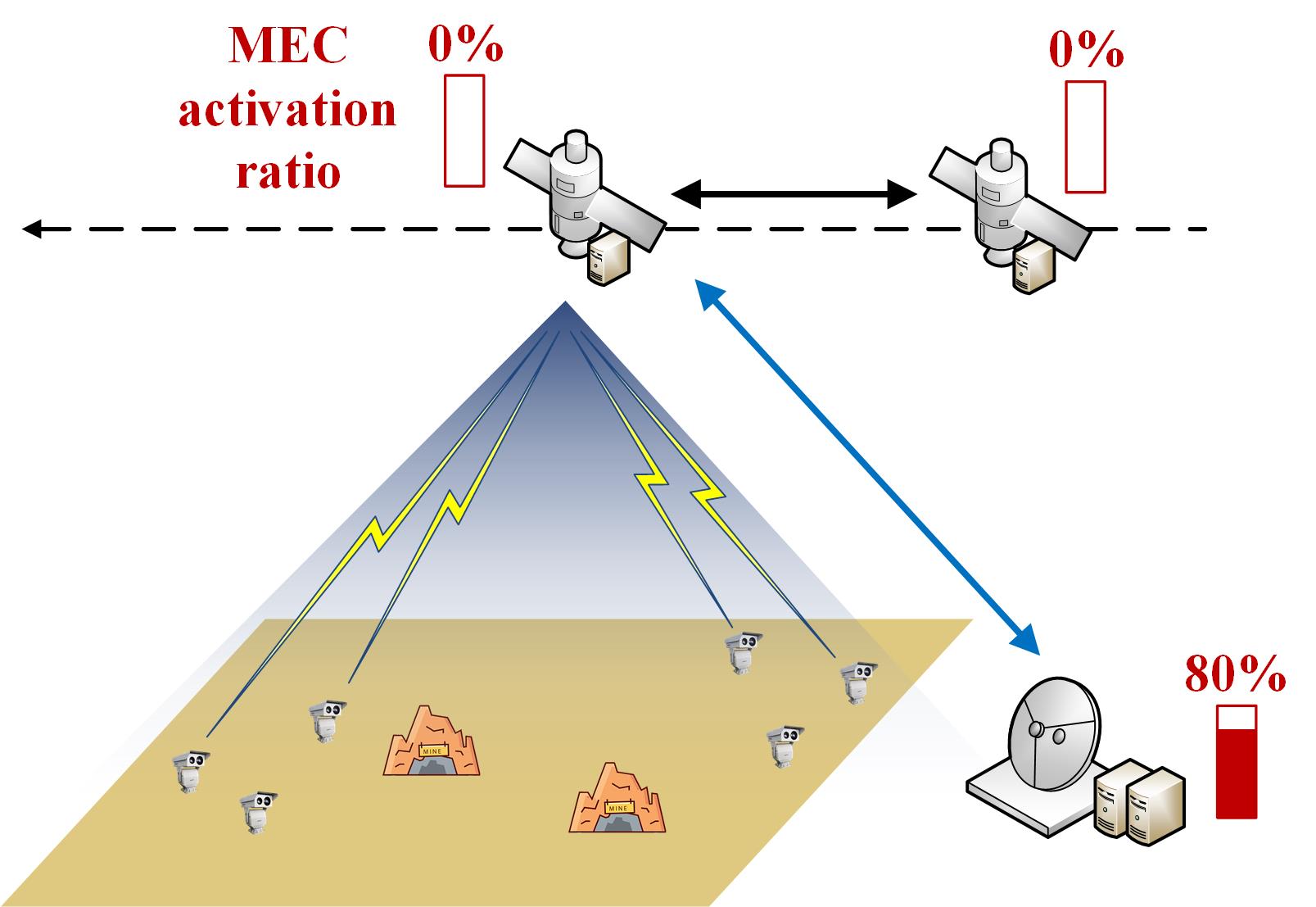}%
		\label{med time 1}}
	\hspace{0.5in}
	\subfloat[]{\includegraphics[width=2.55in]{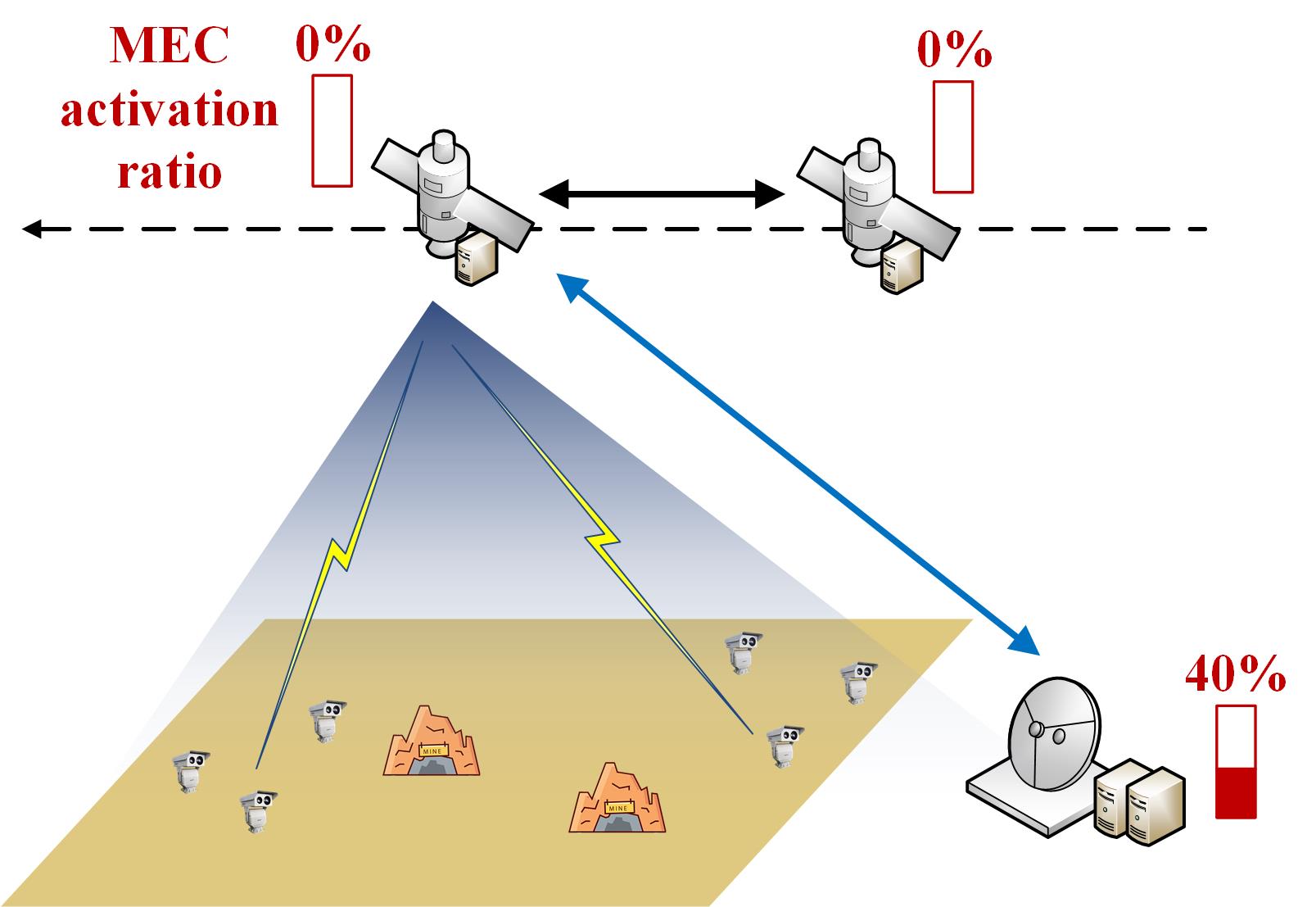}%
		\label{med time 2}}
	\hfill
	\subfloat[]{\includegraphics[width=2.55in]{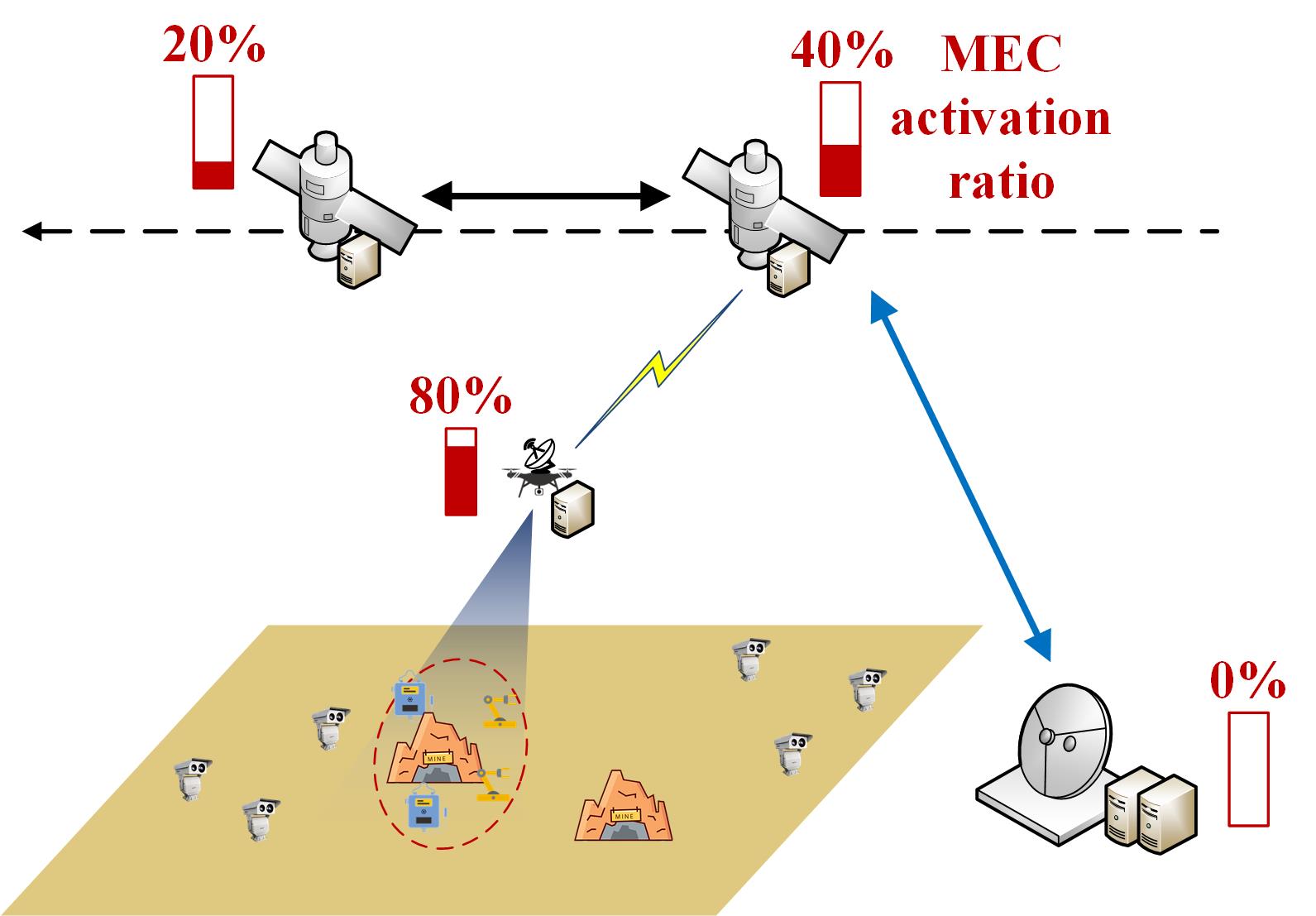}%
		\label{med time 3}}
	\hspace{0.5in}
	\subfloat[]{\includegraphics[width=2.55in]{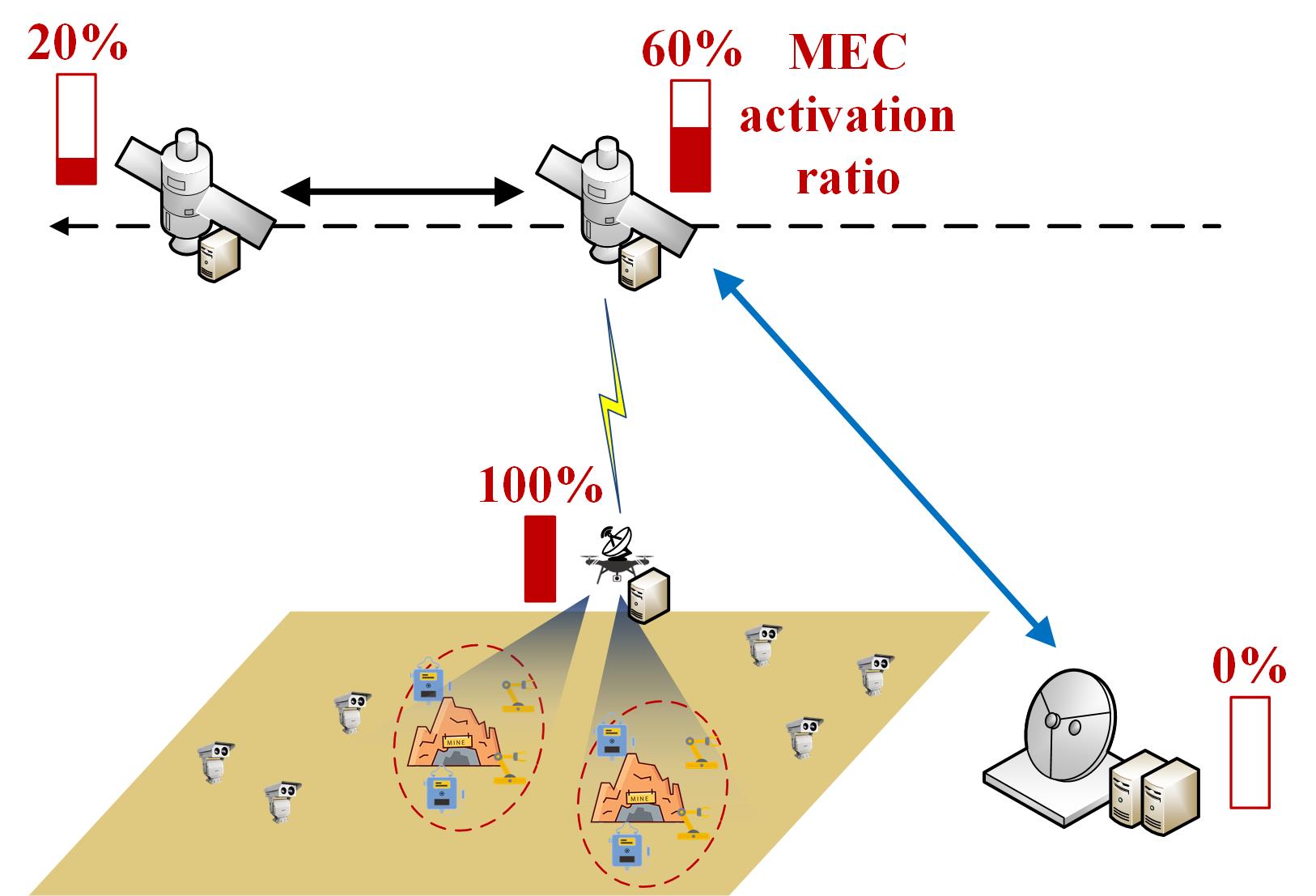}%
		\label{med time 4}}
	\caption{An instance of dynamically adjusting the minimal structure orchestration at a medium timescale to offer everyone-centric customized services.}
\end{figure*}

\begin{figure}[b]
	\centering
	\includegraphics[width=3.2in]{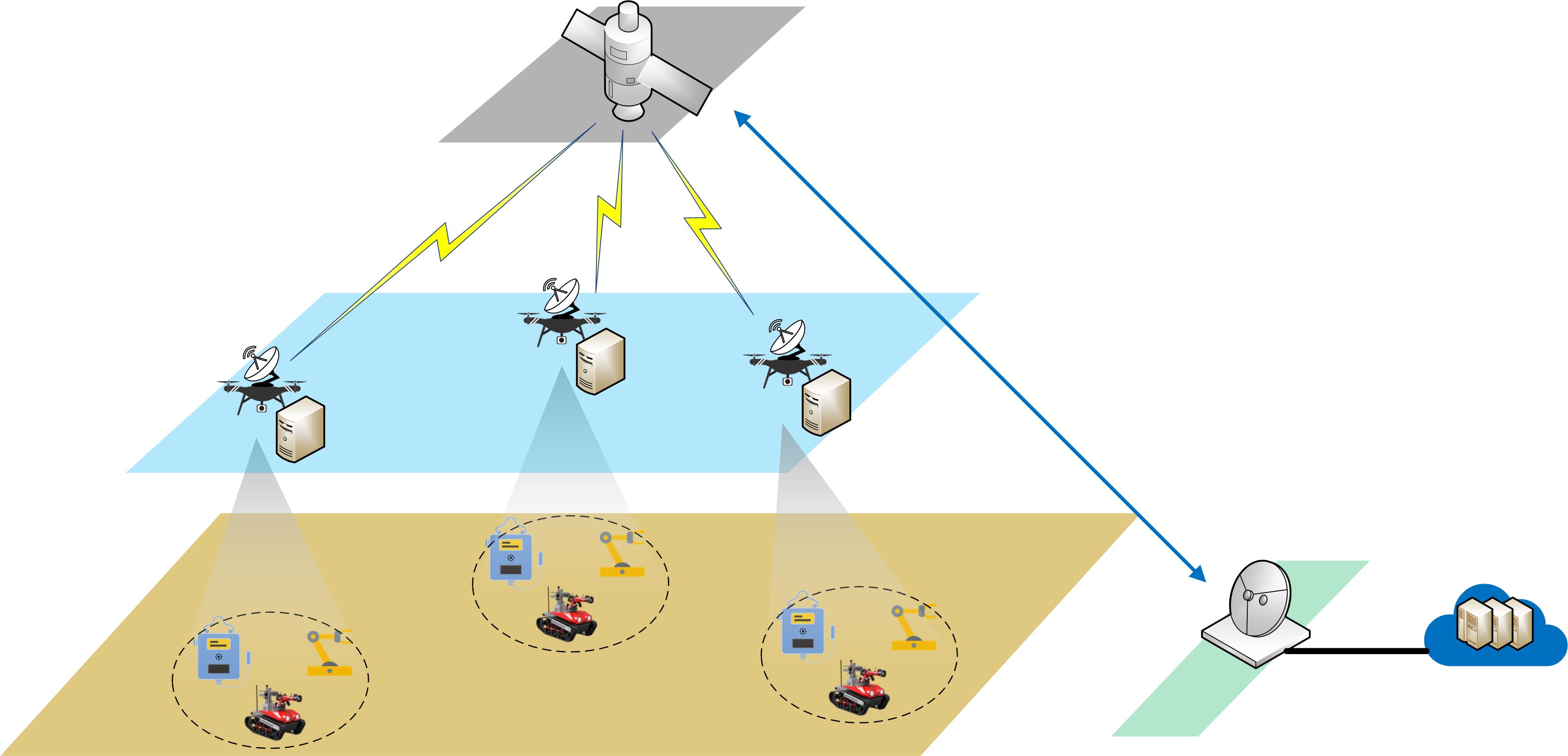}
	\caption{The system model of the case study on process-oriented optimization.}
	\label{case study}
\end{figure}

\section{A Case Study}
In this section, we will provide a case study on process-oriented joint optimization of power allocation and offloading decisions in an extended computing-in-forward-link structure \cite{case study 01}.

As illustrated in Fig. 4, the system consists of a gateway, a satellite, a swarm of UAVs each equipped with an MEC server, a data center and multiple MTDs. This indicates that the system adopts an extended computing-in-forward-link structure with multiple UAVs. Each MTD is associated with the nearest UAV. Additionally, each MTD has one computation task, which can be processed at the UAV server or further uploaded to the data center. The task offloading process is designed under a process-oriented framework. Specifically, the whole process is divided into multiple segmentations. The power allocation and offloading decisions of each MTD are jointly optimized prior to the process to minimize the overall communication and computing latency.

The proposed process-oriented optimization scheme is compared with other schemes. A simple scheme is considered first, where we only use the satellite for communications. Furthermore, a traditional state-oriented scheme is also considered. In contrast to our proposed scheme focused on the whole process, this state-oriented scheme aims to minimize the latency of each state, or in this case, each segmentation. The specifics of the schemes and simulation parameters can be referenced in \cite{case study 01}. From Fig. 5, we can observe that the proposed scheme has better performance compared with the satellite-only scheme. This suggests that the on-demand adjustment of minimal structure orchestration, in this case deploying UAVs with MEC servers, leads to better system performance. Moreover, the proposed process-oriented optimization scheme also outperforms the traditional state-oriented scheme. This suggests that process-oriented optimization can provide a performance gain despite the limited process information.

\begin{figure}[t]
	\centering
	\includegraphics[width=3.3in]{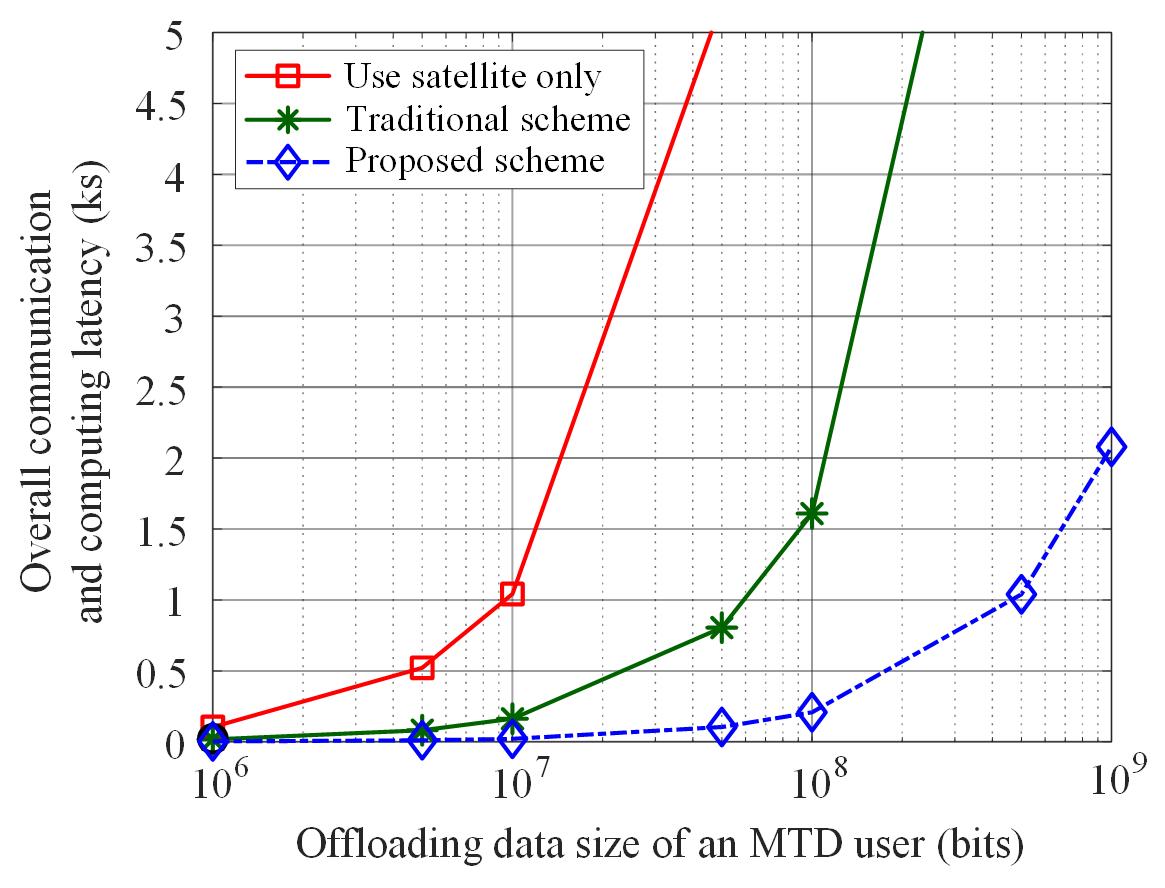}
	\caption{Latency comparison of different schemes.}
	\label{case study sim}
\end{figure}

\section{Open Research Issues}
{\bf Integration with AI:} In the integrated satellite-MEC network, the introduction of the on-demand network orchestration framework raises many problems hard to model. AI-based algorithms can be of great assistance to solving these problems. On the other hand, the integrated satellite-MEC network can support AI-based applications in a ubiquitous and low-latency manner. Therefore, the integration of AI and the network can be an important issue.

{\bf Security:} Security issues require careful consideration in the integrated satellite-MEC network. The inherently large coverage of satellites renders the network susceptible to adverse cyber-attacks. In addition, medium-timescale network adjustments are introduced in the network, which raises new security risks. Therefore, corresponding security measures need to be taken. For instance, blockchain-based methods can be considered to enhance network security.

{\bf Coordination with navigation and sensing:} In addition to communication and computing, the satellite system also has navigation and remote sensing functions. The coordination between the integrated satellite-MEC network and these functions is an interesting problem. Such coordination enables the network to open up new applications that require a comprehensive sensing-communication-computing capability.

{\bf Green network:} Automated vehicles and MEC servers are widely and densely deployed in the integrated satellite-MEC network. This leads to a huge amount of energy consumption and carbon emission. Therefore, having a greener solution is an important issue. Considering the network characteristics, such as dynamic topology, innovative techniques need to be developed to achieve a greener network.

\section{Conclusions}
In this article, we have considered the integration of satellite communications and MEC to efficiently support MTDs in an everyone-centric manner. First, we have provided typical use cases of the integrated satellite-MEC network and the challenges of network design. Then three minimal integrating structures from a systematic perspective have been proposed. To improve resource efficiency, we have established an on-demand network orchestration framework and a process-oriented optimization method. We have then conducted a case study, which shows that medium-timescale network orchestration and process-oriented optimization provide a tremendous performance gain in terms of latency. Finally, we have briefly outlined potential research directions for more intelligent, more secure, and greener satellite-MEC integration.

\vfill


\begin{thebibliography}{1}
\bibliographystyle{IEEEtran}

\bibitem{intro 03}
Y. Yang, M. Ma, H. Wu, {\it et al.}, ``6G network AI architecture for everyone-centric customized services,'' {\it{IEEE Netw.}}, early access, 2022.

\bibitem{intro 01}
X. Fang, W. Feng, T. Wei, Y. Chen, N. Ge, and C.-X. Wang, ``5G embraces satellites for 6G ubiquitous IoT: Basic models for integrated satellite terrestrial networks,'' {\it{IEEE Internet Things J.}}, vol. 8, no. 18, pp. 14399-14417, Sep. 2021.


\bibitem{intro 04}
Y. Yang, ``Multi-tier computing networks for intelligent IoT,'' {\it{Nat. Electron.}}, vol. 2, pp. 4-5, Jan. 2019.



\bibitem{intro 05}
Z. Zhang, W. Zhang, and F. Tseng, ``Satellite mobile edge computing: Improving QoS of high-speed satellite-terrestrial networks using edge computing techniques,'' {\it{IEEE Netw.}}, vol. 33, no. 1, pp. 70-76, Jan./Feb. 2019.

\bibitem{use case 01}
X. Wang, H. Lin, H. Zhang, D. Miao, Q. Miao, and W. Liu, ``Intelligent drone-assisted fault diagnosis for B5G-enabled space-air-ground-space networks,'' {\it{IEEE Trans. Netw. Sci. Eng.}}, vol. 8, no. 4, pp. 2849-2860, Oct.-Dec. 2021.

\bibitem{use case 02}
S. Yu, X. Gong, Q. Shi, X. Wang, and X. Chen, ``EC-SAGINs: Edge-computing-enhanced space–air–ground-integrated networks for Internet of Vehicles,'' {\it{IEEE Internet Things J.}}, vol. 9, no. 8, pp. 5742-5754, Apr. 2022.

\bibitem{challenge 02}
X. You, C.-X. Wang, J. Huang, {\it et al.}, ``Towards 6G wireless communication networks: Vision, enabling technologies, and new paradigm shifts,'' {\it{Sci. China Inf. Sci}}., vol. 64, no. 1, pp. 1–74, Jan. 2021.

\bibitem{challenge 03}
A. D. George and C. M. Wilson, ``Onboard processing with hybrid and reconfigurable computing on small satellites,'' {\it{Proc. IEEE}}, vol. 106, no. 3, pp. 458-470, Mar. 2018.

\bibitem{model 01 018}
C. Ding, J. B. Wang, H. Zhang, M. Lin, and G. Y. Li, ``Joint optimization of transmission and computation resources for satellite and high altitude platform assisted edge computing,'' {\it{IEEE Trans. Wireless Commun.}}, vol. 21, no. 2, pp. 1362-1377, Feb. 2022.

\bibitem{model 02 008}
Z. Song, Y. Hao, Y. Liu, and X. Sun, ``Energy-efficient multiaccess edge computing for terrestrial-satellite Internet of Things,'' {\it{IEEE Internet Things J.}}, vol. 8, no. 18, pp. 14202-14218, Sep. 2021.

\bibitem{model 03 001}
Y. Wang, J. Yang, X. Guo, and Z. Qu, ``A game-theoretic approach to computation offloading in satellite edge computing,'' {\it{IEEE Access}}, vol. 8, pp. 12510-12520, 2020.


\bibitem{model 05 011}
S. Zhang, G. Cui, Y. Long, and W. Wang, ``Joint computing and communication resource allocation for satellite communication networks with edge computing,'' {\it{China Commun.}}, vol. 18, no. 7, pp. 236-252, Jul. 2021.

\bibitem{model 06 066}
G. Cui, X. Li, L. Xu, and W. Wang, ``Latency and energy optimization for MEC enhanced SAT-IoT networks,'' {\it{IEEE Access}}, vol. 8, pp. 55915-55926, 2020.

\bibitem{adjust 01}
X. Li, W. Feng, J. Wang, Y. Chen, N. Ge, and C.-X. Wang, “Enabling 5G on the ocean: A hybrid satellite-UAV-terrestrial network solution,” {\it{IEEE Wireless Commun.}}, vol. 27, no. 6, pp. 116-121, Dec. 2020.

\bibitem{case study 01}
C. Liu, W. Feng, X. Tao, and N. Ge, ``MEC-empowered non-terrestrial network for 6G wide-area time-sensitive Internet of Things,'' {\it{Engineering}}, vol. 8, pp. 96-107, Jan. 2022.


\end{thebibliography}
\end{document}